\documentclass[twocolumn,aps,prl,footinbib,showpacs,
groupedaddress,superscriptaddress]{revtex4-1}

\usepackage{amsmath,amssymb}
\usepackage{graphicx}
\usepackage{dcolumn}
\usepackage{bm}
\usepackage{natbib,amsmath,subfigure,textcomp,verbatim}
\usepackage{hyperref}
\hypersetup{colorlinks=true,citecolor=blue}

\begin{document}

\title{Quantum Frequency Translation of Single-Photon States in Photonic Crystal Fiber}

\author{H.J. McGuinness}
\email{hmcguinn@uoregon.edu}
\author{M.G. Raymer}
\affiliation{Oregon Center for Optics and Department of Physics,
University of Oregon, Eugene, OR USA, 97403 }
\author{C.J. McKinstrie}
\affiliation{Bell Labs, Alcatel-Lucent, Holmdel, New Jersey 07733}
\author{S. Radic}
\affiliation{Department of Electrical and Computer Engineering, University
of California at San Diego, La Jolla, CA 92093}

\date{\today}


\begin{abstract}
We experimentally demonstrate frequency translation of a nonclassical
optical field via the Bragg scattering four-wave mixing process in a
photonic crystal fiber (PCF). The high nonlinearity and the ability to
control dispersion in PCF enable efficient translation between photon
channels within the visible to-near-infrared spectral range, useful in
quantum networks. Heralded single photons at 683 nm were translated to 659
nm with an efficiency of $28.6 \pm 2.2$ percent. Second-order correlation
measurements on the 683-nm and 659-nm fields yielded
$g^{(2)}_{683}(0)=0.21 \pm 0.02$ and $g^{(2)}_{659}(0)=0.19 \pm 0.05$
respectively, showing the nonclassical nature of both fields.
\end{abstract}

\maketitle

\begin{figure}
\includegraphics[width=0.35\textwidth]{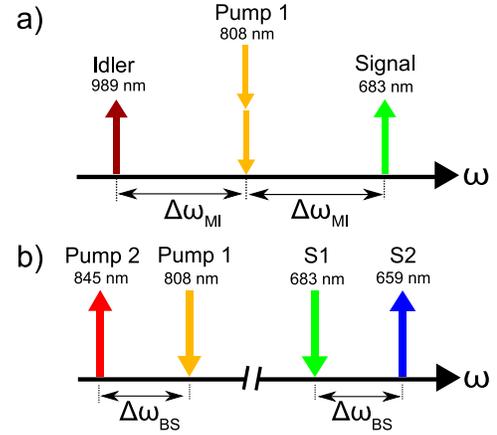}
\vspace{-.05in}
\caption{(color online) a) The modulation interaction process. Two Pump 1 photons
are annihilated while two sideband photons (signal and idler), equally spaced in frequency from
the pump by $\Delta \omega_{\textrm{MI}}$, are created. Up(down) arrows indicate creation(destruction).
b) The Bragg-scattering quantum frequency translation process. Photons from
Pump 1 and the S1 mode are annihilated while photons
in Pump 2 and the $s2$ mode are created. The frequency separation between the
pumps $\Delta \omega_{\textrm{BS}}$ is equal to the separation S1 and $s2$.}
\label{fig:FWM}
\end{figure}

As more advanced quantum-information applications and systems are created,
it is likely that the sharing of quantum information between remote
devices will be desirable, and a quantum network will be needed
\cite{Cirac}. A good candidate to transfer such information is the single
photon, which is relatively robust against loss or decoherence, allowing
transfer of entanglement between remote locations. Such photons can travel
long distances through optical fibers, which function optimally in
particular wavelength (``telecom") ranges. Quantum frequency translation
(QFT), in which a photon at one central frequency is annihilated and
another photon at a different central frequency is created (see Figure
\ref{fig:FWM}b)), serves three important purposes in quantum networks--It
allows quantum devices (memories and processors) that operate at different
optical frequencies to communicate via a quantum channel; It allows
low-loss, long-distance exchange over fiber between two quantum devices
that operate at frequencies other than the telecom ones; It allows
converting photons to frequencies where the optimal detectors operate. 
To be useful in quantum networks, QFT must 1) allow flexible choices of
photon frequencies within the visible and near IR, 2) preserve the nature
of the original state other than its central frequency, including any
entanglement with other systems, and 3) must not introduce additional
unwanted ``noise" photons. QFT in optical fiber is predicted to satisfy
all of these requirements \cite{McKinstrie2}. We present the first
demonstration of QFT in optical fiber, and show that it preserves the
nonclassical nature of the single-photon field being translated.

Frequency conversion in second-order $\chi^{(2)}$ nonlinear optical media
such as crystals via three-wave mixing has been studied in much depth,
especially for coherent-state fields \cite{Boyd}. Sum frequency generation 
allows a weak field, when combined with a strong pump field, to be
translated to a higher frequency (upconversion). Difference frequency
generation allows for the creation of the conjugate of a weak field, in
the presence of a pump, which can either be at a higher or lower
(downconversion) frequency than the original field. At the single-photon
level, it has been shown that nonclassical intensity correlation between
two 1064-nm fields was maintained when one beam was upconverted to 532 nm
\cite{Huang}. Also, time-energy entanglement between two fields at 1555 nm
and 1312 nm was shown to be preserved after the 1312-nm field was
upconverted to 712 nm \cite{Tanzilli}. Most recently, a single-photon
state at 1300 nm was shown to maintain its nonclassical nature after
upconversion to 710 nm in a periodically-poled $\textrm{LiNbO}_3$
quasi-phase-matched waveguide \cite{Rakher}.

A disadvantage of  $\chi^{(2)}$-based QFT is that the difference between
the two frequencies involved must equal the frequency of a strong pump
field, making it impractical to translate between two closely spaced
channels, for example within the red portion of the spectrum. In contrast,
$\chi^{(3)}$-based QFT, which we use here, requires only that two pump
fields have a frequency difference equal to that by which one aims to
translate the quantum state of interest. This, combined with flexible
phase matching provided by photonic crystal fiber (PCF) as the medium,
enables QFT within the visible-to-near-IR range, with no lower bound to
the frequency separation of channels.

The process we developed for noiseless QFT is two-pump Bragg-scattering
(BS), a nondegenerate, four-wave mixing (FWM) process analogous to sum- or
difference-frequency generation in $\chi^(3)$ media. As opposed to other
fiber-based frequency conversion processes which involve amplification and
its attendant spontaneous-emission noise, BS does not amplify the 
translated field, and is therefore theoretically noiseless
\cite{McKinstrie3}. In this sense it is also analogous to a passive beam
splitter, meaning that analogues of two-photon interference apply.
\cite{Raymer2} Furthermore, it has been proved that BS is capable of
translating arbitrary multi-photon states of one spectral mode to another
spectral mode \cite{McKinstrie3}. For classical fields, BS was first shown
at telecom wavelengths \cite{Inoue}. Recently, several experiments have
explored the BS process in achieving near-unity conversion efficiency,
low-noise wavelength translation from 1545 nm to 1365 nm
\cite{Uesaka,Gnauck,Mechin}.

The $\chi^{(3)}$ BS process in optical fiber has additional advantages
compared with sum/difference-frequency conversion in bulk and quasi-phase
matched $\chi^{(2)}$ materials. Since the process occurs in optical fiber, 
it is straightforward to couple the translated field into another fiber
with small loss. This makes translation in fiber well suited for use in
quantum networks, since such networks would be sensitive to small losses.
Often, the process is carried out in a single-mode fiber, where both input
and output are in well defined spatial modes, which is not always the case
for sum/difference generation materials. This allows the output to be more
easily integrable with devices or experiments. Also, since the efficiency
of the process is proportional to the product of the input power of each
pump, one pump can be weak if the other pump compensate by being strong.
We emphasize that PCF offers unprecedented control over the medium's
dispersive properties, allowing translation to occur anywhere within the
visible-to-infrared by engineering the fiber in a particular, controllable
manner \cite{Birks}. With the advent of hollow-core PCF \cite{Cregan},
media other than silica maybe used for the BS interactions, adding the
possibility of using more highly nonlinear materials than glass and
gaining further control over dispersion of the fiber.

In this Letter we report for the first time, to our knowledge, frequency
translation in optical fiber carried out with verifiably nonclassical
fields. Also, to the best of our knowledge, this is the first time BS
translation in a $\chi^{(3)}$ medium has been reported in the visible
regime. Briefly, a single-photon wave packet with a central wavelength 683
nm is created in a custom made PCF, which is then coupled into another PCF
fiber along with two pump fields. The BS process takes place and with some
probability (efficiency) the single-photon wave packet is translated to
another wave packet with a central wavelength 659 nm (otherwise it is left
alone). The translated and untranslated channels are then monitored with
single-photon detectors which measure the degree of nonclassicality and
the overall efficiency of the BS process.

\begin{figure}
\includegraphics[width=000.48\textwidth]{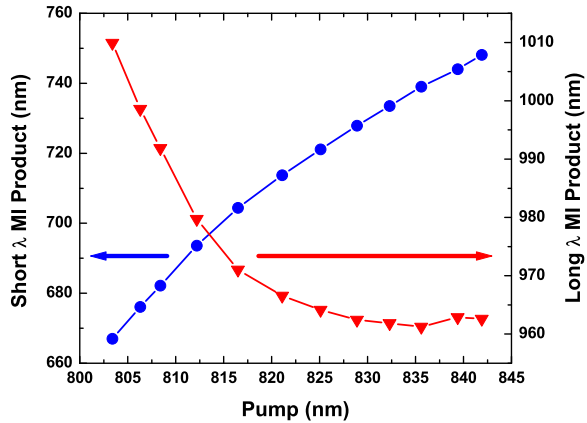}
\vspace{-.4in}
\caption{(color online) Experimental data concerning the MI process
in Fiber 1. The short (blue circle) and long (red triangle) wavelength
sidebands refer to the signal and idler fields, respectively.}
\label{fig:MIdata}
\end{figure}

The single-photon states that were translated were produced in a fiber via
the one-pump FWM process called ``modulation instability" (MI), shown in
Figure \ref{fig:FWM}a). \cite{McKinstrie,Fulconis:05} In MI, two pump
($p$) photons are annihilated in the medium and two non-degenerate photons
are spontaneously created. The central wavelengths of the produced
``sideband" photons, also called signal (short wavelength, $s$) and idler
(long wavelength, $i$), are determined by energy conservation
($2\omega_p=\omega_s +\omega_i$) and phase-matching requirements
($2\beta_p = \beta_s + \beta_i$) \cite{Agrawal}. With a single CW laser
input at power $P$, fiber nonlinearity $\gamma$ and fiber length $L$,
where $P\gamma L \ll 1$, the state produced is given by
\begin{align}\label{CWstate}
| \psi \rangle \approx |0_s,0_i\rangle  + \epsilon|1_s,1_i\rangle
+ \epsilon^2|2_s,2_i\rangle + \dots
\end{align}
where $\epsilon \ll 1$ is a function of $P,\gamma$ and $L$. Since
$\epsilon$ is small the state is mostly vacuum, but if the idler (signal)
channel is incident upon a detector and the detector registers an event,
it is very likely that the signal (idler) is in the $|1\rangle$ Fock
state. This process is called heralding the signal (idler). For pulsed
pump input the state is more complicated, but is not crucial for
understanding the relevant physics in this Letter \cite{Chen,Garay}. Phase
matching in the PCF that we used in this experiment allows continuous
tuning of the sidebands over a wide range (Figure \ref{fig:MIdata}).

After single-photon wave packets are created, one is sent into a second
fiber for QFT via BS \cite{McKinstrie}. Two pump fields $p1$ and $p2$ and
the signal field $s1$ are coupled into the fiber, leading to the
annihilation of the signal field and creation of a translated signal field
$s2$, as shown in Figure \ref{fig:FWM}b). The fields must obey energy
conservation ($\omega_{p1} +\omega_{s1}=\omega_{s2} +\omega_{p2}$) and be
phase-matched ($\beta_{p1} +\beta_{s1} = \beta_{s2} + \beta_{p2}$) for
efficient translation to occur. Treating the pumps as classical fields and
the signals as quantum fields, the process is governed by the Hamiltonian
\cite{McKinstrie2}
\begin{align}\label{Ham}
H = \delta(a^{\dagger}_{s1} a^{}_{s1} + a^{\dagger}_{s2} a^{}_{s2})
 + \kappa a^{\dagger}_{s1} a^{}_{s2} + \kappa^{*}a^{\dagger}_{s2} a^{}_{s1}
\end{align}
where $a^{\dagger}$ and $a$ are creation and annihilation operators. The
quantities $\delta$ and $\kappa$ relate to the dispersion mismatch between
pumps and sidebands, and the effective nonlinearity, respectively.
Utilizing the spatial equations-of-motion $d_z a_j=i[a_j,H]$, the
operators can be solved as functions of position along the fiber, yielding
\begin{align}\label{Ham}
a^{}_{s1}(z) & =  \mu(z)a^{}_{s1}(0) +\nu(z)a^{}_{s2}(0) \\
a^{}_{s2}(z) & =  -\nu^*(z)a^{}_{s1}(0) +\mu^*(z)a^{}_{s2}(0),
\end{align}
where the transfer functions $\mu$ and $\nu$ are
\begin{align}\label{Ham}
\mu(z) & =  \textrm{cos}(kz) + i\delta \textrm{sin}(kz)/k \\
\nu(z) & =  i\kappa \textrm{sin}(kz)/k,
\end{align}
and where $k=(|\kappa|^2 + \delta^2)^{1/2}$ and $|\mu|^2 +|\nu|^2=1$. The
process conserves total photon number, like a beam splitter. If the
parameters are set such that $|\nu(L)|=1$, then all of the signal field
will be translated to the idler field by the end of the fiber.

The above analysis is valid for single-mode, CW fields. For pulsed fields
a more general theory is appropriate \cite{Raymer2}. The temporal and
spectral properties of all input fields play a significant role in
determining the nature of the translated field. For efficient translation,
temporal overlap of the fields must be maximized and care must be taken
that the bandwidth of the input field is not greater than the bandwidth of
the BS process, which is a function of fiber dispersion, length and pump
bandwidth.

In order to demonstrate that the $s1$ input and the $s2$ output are
nonclassical, the conditional second-order degree of coherence
$g^{(2)}(0)$ is measured for these fields. A result of $g^{(2)}(0)<1$
indicates a nonclassical field \cite{Loudon}. For a heralded field,
$g^{(2)}(0)$ is given experimentally by
\begin{align}\label{g2.3D}
g^{(2)}(0) = \frac{N_{ABC} N_C}{N_{AC}N_{BC}}
\end{align}
where $A$ and $B$ label the detectors monitoring the output of the
heralded signal field, and $C$ labels the heralding detector in the idler
field. The quantities $N_{i},N_{ij},N_{ijk}$ are the number of single,
double coincidence, triple coincidence events between detectors $i,j$ and
$k$ over the time interval of the data collection, respectively
\cite{Beck}.

\begin{figure}
\includegraphics[width=0.48\textwidth]{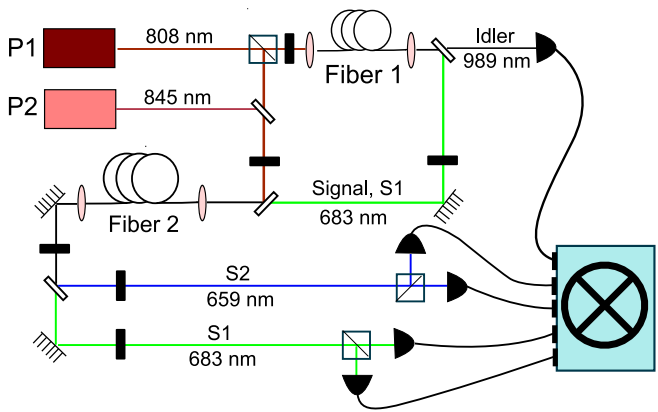}
\vspace{-0.02\textwidth}
\caption{(color online) Setup used to create and frequency translate single-photon
states. Both pumps $p1$ and $p2$ are 100 ps titanium sapphire lasers. A single photon
is created in Fiber one at 683-nm, heralded, and combined with $p1$ and $p2$ to be
translated to 659-nm in Fiber two. Five detectors measure $g^{(2)}(0)$ of both
683-nm and 659-nm channels, with the 989-nm channel detector acting as the herald.
Opaque rectangles represent either longpass or bandpass filters and transparent
rectangles signify dichroic mirrors.}
\label{fig:setup}
\end{figure}

The experimental apparatus is shown in Figure \ref{fig:setup}. Pump 1 and
Pump 2 were 100 ps titanium-sapphire lasers operating at 808 nm and 845
nm, respectively. Both lasers, having repetition rates of 76 MHz, were
long-pass filtered to extinguish any light in the quantum channels. A
small percentage of Pump 1 was split off and sent to 32 m of a custom-made
solid-core PCF, denoted as Fiber 1, having a zero-dispersion wavelength 
(ZDW) at approximately 796 nm. Sidebands were created at 683 nm and 989 nm
using the MI process in which phase matching was achieved by using the
fiber birefringence \cite{Fan,Smith}. These beams were split, with the
989-nm idler being sent to a single-photon detector for heralding. Both
sidebands were filtered with 13-nm bandpass filters to remove unwanted
spontaneous Raman scattering.

The rest of Pump 1, and all of Pump 2, were combined on a dichroic mirror,
with the resulting beam further combined with the 683-nm signal from Fiber
1 by use of another dichroic mirror.  The pulses of both pumps and the
683-nm signal pulse were overlapped temporally. All beams had the same
linear polarization. All three beams were incident upon 20 meters of Fiber
2 (Crystal Fibre, model number NL-PM-750), and aligned on one of the
principal birefringence axes of this fiber. Around 20(30) mW from Pump
1(2) was coupled into Fiber 2. We achieved over 31 percent coupling of the
683-nm beam from Fiber 1 into Fiber 2 prior to translation. The state of
the 683-nm field was frequency translated with a certain probability to
the field centered around 659 nm. After exiting Fiber 2, the pump
components were filtered out of the beam, which was then separated into
the 683-nm and 659-nm channels. Both channels were filtered with 13-nm
bandpass filters. Conditional second-order correlation measurements
($g^{(2)}(0)$) of the 683-nm and 659-nm channels were carried out in
coincidence with the 989-nm MI idler channel.

To verify that the translation process is occurring with high efficiency,
we measured both the depletion efficiency of the 683-nm $s1$ input and the
creation efficiency of the 659-nm $s2$ output, which should be equal. The
depletion efficiency was obtained by measuring the rates of counts (minus
pump noise counts) in the output 683-nm $s1$ channel with and without the
pump beams coupled into Fiber 2. The depletion efficiency equals one minus
the ratio of these two count rates. The result, $28.6 \pm 2.2$ percent, is
not dependent on comparing quantum efficiencies of two detectors. To
measure the creation efficiency of the 659-nm $s2$ channel, the count rate
in this channel was monitored with and without the pumps. When detector
efficiencies are taken into account, this measurement yields $29.4 \pm
2.4$ percent, agreeing with the depletion measurement to within error.

To show the nonclassical nature of the translated and untranslated light,
$g^{(2)}(0)$ measurements were performed on both the 683-nm and 659-nm
channels at the output of the fiber, yielding $0.21 \pm 0.02$ and $0.19
\pm 0.05$, respectively. To determine the average value and error of the
$g^{(2)}(0)$ measurements, 30 runs of the experiment lasting 20 seconds
each were performed.

There are several factors which cause the measured values of $g^{(2)}(0)$
to be nonzero. After heralding, the photon state to be translated will be
predominantly the $|1\rangle$ Fock state. But the state will also include
small amplitudes for number states above 1, which will increase the value
of $g^{(2)}(0)$. For states generated via heralded MI there will always be
a trade-off between higher count rates and high single-photon state
probability (unless a number-resolving detector is used for heralding).
Low detector efficiency (12 percent) for the 989-nm channel demanded
higher input power than desired. Pump 1 input power in Fiber 1 was chosen
to give reasonably high photon flux and coincident count rates while
guaranteeing the heralded state \eqref{CWstate} was predominantly the
$|1\rangle$ state. Another possible cause for increased $g^{(2)}(0)$ was
from accidental coincidence counts from Raman scattering, detector dark
counts and other noise. Over a given time, an experiment that counted
$N_i$ idler counts and $N_s$ signal counts, both which derived from $N_p$
pulses, would expect to register $N_i N_s/N_p$ accidental coincidences if
the idler and signal beams were independent and poisson. Therefore any
measurement that depends on coincidence counts from a correlated source,
such as sideband creation via MI, needs to register coincidences above
this value. In our experiment the coincidence level was approximately 8.2
times higher than the expected accidental coincidence value for
coincidences between the 683-nm channel and the 989-nm heralding channel.
The coincidence level for the 659-nm translated channel was approximately
6.5 times higher than the expected accidental coincidences.

These same noise counts affected the measured efficiency of the
translation process. Approximately 24 percent of counts in the 659-nm
channel, and 11 percent in the 683-nm channel, were from pump noise (e.g.,
Raman scattering). Most of the noise originated from the 808-nm pump. It
is clear how such noise could be drastically reduced. Due to the
wavelength range of the available pumps, and the dispersion
characteristics of the fibers, it was necessary for Pump 1 to operate at
808 nm, which is only 50 to 70 nm away from Fiber 2's ZDW. Many nonlinear
effects grow exponentially stronger the closer the pump field wavelength
is to the fiber's ZDW. Operating the pump farther from the ZDW, and
farther from the signal and idler modes in general, would strongly
decrease the noise in these channels. Also, the Stokes and anti-Stokes
Raman noise can be strongly reduced by cooling Fiber 2 to liquid nitrogen
temperature \cite{Takesue}.

Theory predicts that BS is capable of 100 percent translation efficiency
of arbitrary states \cite{McKinstrie3}. In our experiment various factors
limit the efficiency. The most likely factor was the relatively large
spectral width of the 683-nm field created by MI. This conjecture is
supported by the fact that the full width at half maximum (FWHM)of the
output translated field was less than that of the input untranslated
state, indicating the translation process had a narrower acceptance
bandwidth than the bandwidth of the input field. These widths were
measured using an input signal at 683-nm created by high-gain MI, with a
FWHM of approximately 2.0 nm, while the translated field had a FWHM equal
or less than 1.5 nm, the minimum width resolution of the spectrometer. The
wide FWHM of the input field was due to Pump 1 being close in wavelength
to the ZDW of Fiber 1 \cite{Marhic, Garay}. For this experiment it was
unfortunately necessary for the pump and ZDW of Fiber 1 to be close in
wavelength, but this is not indicative of a fundamental limitation of the
BS translation process.

We have demonstrated frequency translation of nonclassical states of light
via the four-wave mixing Bragg scattering process. This is the first time
BS on the quantum level in a $\chi^{(3)}$ media has been reported. This is
also the first time translation within the visible regime has been
reported. While translation in $\chi^{(3)}$ media requires two pumps, as
opposed to translation in $\chi^{(2)}$ media, which requires only one,
there are a number of advantages such a method offers. Because the
translating device could be easily coupled to an optical fiber network, it
is the clear choice for translation in a quantum network. The two-pump
configuration allows for a more flexible achievement of phase-matching
conditions, allowing translating between spectrally close channels. This
in turn opens possibilities for two-photon (Hong-Ou-Mandel) interference
between photons of different colors \cite{Raymer2} and ultimately
linear-optics quantum computing \cite{Knill} using multiple frequency
channels.

We thank S. van Enk for helpful comments. This work was supported by NSF
Grant ECCS-0802109.

\bibliography{QuantBS}

\end{document}